\documentclass[conference]{IEEEtran}
\IEEEoverridecommandlockouts
\usepackage{cite}
\usepackage{amsmath,amssymb,amsfonts}
\usepackage{algorithmic}
\usepackage{graphicx}
\usepackage{textcomp}
\usepackage{xcolor}
\usepackage{multirow}
\usepackage{hyperref}% Por mi
\usepackage{array}% Por mi
\usepackage{makecell}
\newcolumntype{C}[1]{>{\centering\arraybackslash}m{#1}}% Por mi
\newcolumntype{L}[1]{>{\raggedright\arraybackslash}m{#1}}% Por mi
\newcolumntype{J}[1]{>{\arraybackslash}m{#1}}% Por mi
\def\BibTeX{{\rm B\kern-.05em{\sc i\kern-.025em b}\kern-.08em
    T\kern-.1667em\lower.7ex\hbox{E}\kern-.125emX}}
\begin{document}

\title{Overview of Risk Assessment and Management for Intelligent Systems under the AI Act and Beyond\\
%ARES: A Platform for Evaluating Adaptive Role-Based Social Engineering Risks in Human–AI Games
\thanks{This research was supported by Cátedra ENIA UAM-VERIDAS en IA Responsable (NextGenerationEU PRTR TSI-100927-2023-2), M2RAI (PID2024-160053OB-I00, MICIU/FEDER), TRUST-ID (PID2025-173396OB-I00, MICIU/AEI and the EU) and PowerAI+ (SI4/PJI/2024-00062, Comunidad de Madrid and UAM). Javier Irigoyen is supported by an FPI fellowship from MINECO/FEDER.}}
\author{
\IEEEauthorblockN{
Javier Irigoyen\IEEEauthorrefmark{1},
Roberto Daza\IEEEauthorrefmark{1}\IEEEauthorrefmark{2},
Aythami Morales\IEEEauthorrefmark{1}\IEEEauthorrefmark{3},
Julian Fierrez\IEEEauthorrefmark{1},
Ruben Tolosana\IEEEauthorrefmark{1},\\
Ruben Vera-Rodriguez\IEEEauthorrefmark{1},
Francisco Jurado\IEEEauthorrefmark{2}, and
Alvaro Ortigosa\IEEEauthorrefmark{2}
}

\IEEEauthorblockA{
\IEEEauthorrefmark{1}\textit{BiometricsAI, Universidad Autónoma de Madrid (UAM), Spain}
}

\IEEEauthorblockA{
\IEEEauthorrefmark{2}\textit{GHIA, Universidad Autónoma de Madrid (UAM), Spain}
}

\IEEEauthorblockA{
\IEEEauthorrefmark{3}\textit{Universidad de Las Palmas de Gran Canaria (ULPGC), Spain}
}

\IEEEauthorblockA{
Corresponding author: javier.irigoyen@uam.es
}
}

% {\footnotesize \textsuperscript{*}Note: Sub-titles are not captured for https://ieeexplore.ieee.org  and

% \and
% \and
% \and
% \and
% \and

\maketitle

\begin{abstract}

The society and emerging risk-based regulatory frameworks for AI underscore the need for rigorous risk assessment to ensure safe and reliable AI systems. In response to this imperative, this paper presents an overview of AI risk assessment (identification and analysis) and management methodologies. It begins by reviewing the worldwide regulatory landscape that drives the need for systematic AI risk assessment. Then we characterize the spectrum of AI-related risks identified in the literature, from technical failures to ethical and social impacts. Subsequently, it reviews key risk assessment methodologies proposed for AI systems, focusing on general frameworks. The paper highlights best practices and illuminates methodological gaps, highlighting areas for further research on AI risk assessment. 

\end{abstract}


\section{Introduction}
\label{intro}

The proliferation and integration of Artificial Intelligence (AI) technologies in multiple sectors has led to significant innovations and improvements in efficiency, decision-making, and automation. Despite these advantages, the widespread use of AI also introduces considerable risks, such as biased outcomes, privacy violations, lack of transparency, and cybersecurity vulnerabilities. As AI systems become increasingly autonomous, these risks pose significant legal, ethical, and operational challenges for organizations and society at large. Consequently, there is a growing emphasis from regulatory bodies, policymakers, and academia on systematically assessing and managing AI-associated risks, exemplified by the recent entry into force of the European Union's Artificial Intelligence Act (AI Act).

The AI Act represents a landmark regulatory approach that aims to create a standardized framework to identify and mitigate risks posed by AI systems. Its risk-based approach categorizes AI applications according to potential harm, mandating stringent assessment requirements for high-risk AI applications. Although such regulations provide a foundational structure for risk governance, there remains a need for practical methodologies to implement these regulatory requirements effectively within organizations.

Recent academic literature has offered information on methodologies and frameworks for AI risk assessment, including contributions on interpretability, fairness, robustness, and sustainability. Despite these advances, current approaches often remain theoretical or domain-specific, lacking comprehensive empirical validation in diverse organizational contexts. Furthermore, existing research frequently addresses individual risk dimensions in isolation or lacks integration into a coherent, universally applicable framework.

Our paper addresses these critical gaps by systematically reviewing existing AI risk assessment approaches, synthesizing insights from the literature, regulatory requirements, and practical implementations within organizations. Through our comprehensive review, our aim is to provide clarity on best practices and shortcomings of current methodologies, ultimately contributing to the development of robust, empirically validated guidelines for effective AI risk assessment and compliance under the new regulatory landscape.

%La regulación actual empuja a desarrollar una IA más ética y segura. Se establece como requisito desarrollar nuevos stándares. Los existentes son muy generales. Hace falta un trabajo previo para entender bien como definir los riesgos de la IA y como medirlos. En este trabajo, tratamos de abordar ese gap entre lo que pide la sociedad y la regulación y los estándares que tienen que aparecer en un futuro próximo. Main research questions:
%\begin{itemize}
%    \item How can we define risks in AI?
%    \item How can we measure risks in AI? Aproximaciones basadas en la tecnología (red neuronal, SVM, etc...) o la aplicación?
%    \item How can we select the best AI risk assessment approach: general assessment, application-dependent?
%\end{itemize}

%% Labels are used to cross-reference an item using \ref command.

\section{Introduction to Ethical Guidelines and Trustworthy AI in the European Context}

The European Commission's work on Artificial Intelligence (AI) governance has been guided by the High-Level Expert Group on Artificial Intelligence (AI HLEG). Its \textit{Ethics Guidelines for Trustworthy AI} \cite{cannarsa2021ethics}, published in April 2019, set foundational principles for responsible AI deployment in Europe and introduced Trustworthy AI as a governance ideal built on three interconnected principles.

\subsection{General Principles of Trustworthy AI}

The principles of Trustworthy AI \cite{2023_SNCS_Human-Centric_Pena,diaz2023connecting} can be summarized in three main pillars:

\begin{enumerate}
    \item \textbf{Lawful:} AI systems must comply with the legal frameworks, regulations, and obligations applicable in the European Union.

    \item \textbf{Ethical:} Beyond legal compliance, AI must respect fundamental ethical principles and societal values, prioritizing human dignity, individual autonomy, and fairness so that technology serves rather than undermines society.

    \item \textbf{Robust:} AI must be technically resilient, maintaining accuracy, reliability, security, and consistency while accounting for its broader social context to minimize risks and unintended harm.
\end{enumerate}

Together, these principles define the baseline that an AI system must meet to be considered trustworthy in the European perspective outlined by the AI HLEG.

\subsection{Operationalization of Principles into Specific Requirements}

The AI HLEG translated these broad principles into seven practical requirements that systems should satisfy throughout their life cycle:

\begin{enumerate}
    \item \textbf{Human Agency and Oversight:} AI systems should empower users, support informed decision-making, and protect fundamental rights. Oversight mechanisms—ranging from "human-in-the-loop" to "human-in-command" approaches—must ensure that AI augments rather than overrides human judgment.

    \item \textbf{Technical Robustness and Safety:} AI must be secure, resilient, accurate, reliable, and reproducible, with safety mechanisms and follow-up plans to limit harm from unexpected failures or errors.

    \item \textbf{Privacy and Data Governance:} AI systems must respect privacy and follow robust data governance practices, safeguarding data quality, integrity, and accessibility while complying with data protection rules such as the GDPR.

    \item \textbf{Transparency:} Clear information on AI capabilities, limitations, logic, and decisions must be provided. Stakeholders should know when they interact with AI and understand AI-generated outcomes through traceability and explainability.

    \item \textbf{Diversity, Non-discrimination, and Fairness:} AI systems must avoid biases that could marginalize or unfairly disadvantage individuals or vulnerable groups. Fairness, inclusivity, and accessibility should guide development and deployment.

    \item \textbf{Societal and Environmental Well-being:} The broader impacts of AI on society and the environment must be carefully managed. AI technologies should foster sustainability, mitigate environmental harm, and consider the long-term welfare of future generations.

    \item \textbf{Accountability:} Clear and enforceable mechanisms must be established to attribute responsibility for AI systems' actions and outcomes. Auditability, traceability, and effective redress procedures ensure accountability, particularly for critical or high-risk applications.
\end{enumerate}

These requirements translate lawful, ethical, and robust AI into practical guidance for industry, policymakers, and regulators. Later EU policy, especially the proposed \textbf{Artificial Intelligence Act (AI Act)}, builds on them by formalizing a legally enforceable framework for managing AI risks.

\section{Towards Risk-based AI Systems}

Building on the ethical framework provided by the AI HLEG, the European Commission proposed the \textbf{Artificial Intelligence Act (AI Act)} to regulate AI systems systematically. The Act adopts a \textit{risk-based approach}, classifying AI applications into four categories:

\begin{enumerate}
    \item \textbf{Unacceptable Risk:} Applications incompatible with EU values, such as social scoring or manipulative systems influencing vulnerable individuals, are prohibited.
    
    \item \textbf{High Risk:} Systems whose malfunction could affect safety, fundamental rights, or well-being—such as medical diagnostics, biometric identification, or critical infrastructure management—are subject to stringent obligations.

    \item \textbf{Limited Risk:} Systems that require transparency obligations but do not pose severe threats, such as chatbots or deepfakes, must inform users that they are interacting with AI.

    \item \textbf{Minimal or No Risk:} Applications such as spam filters or video game AI, which present negligible risk, face minimal or no obligations.
\end{enumerate}

For \textbf{High Risk AI systems}, the AI Act defines obligations closely aligned with Trustworthy AI requirements, including:

\begin{itemize}
    \item Data and data governance.
    \item Risk management system.
    \item Technical documentation and record keeping.
    \item Transparency and provision of information to users.
    \item Human oversight.
    \item Accuracy, robustness, and cybersecurity.
    \item Quality management system.
\end{itemize}

The risk management system obligation can be detailed into the following requirements:

\begin{enumerate}
    \item \textbf{Risk management system:}
    \begin{itemize}
        \item Characterization of the AI system.
        \item Continuous operation throughout the life cycle.
    \end{itemize}

    \item \textbf{Risk management process:}
    \begin{itemize}
        \item Identification of risks affecting the AI system.
        \item Estimation and evaluation of risks caused by reasonably foreseeable misuse.
        \item Evaluation of other possible risks.
    \end{itemize}

    \item \textbf{Risk management measures to eliminate or reduce risks:}
    \begin{itemize}
        \item Adequate design and development.
        \item Mitigation and control measures.
        \item Provision of information to users.
        \item User training.
    \end{itemize}

    \item \textbf{Conditions required to operate the AI system:}
    \begin{itemize}
        \item User capacities (e.g., technical knowledge, experience, education, training).
        \item Configuration of the intended operating environment.
    \end{itemize}

    \item \textbf{Testing of the AI system:}
    \begin{itemize}
        \item System performance.
        \item System compliance with prior requirements.
    \end{itemize}
\end{enumerate}

\begin{figure*}[t]
  \centering
  \includegraphics[width=.7\linewidth]{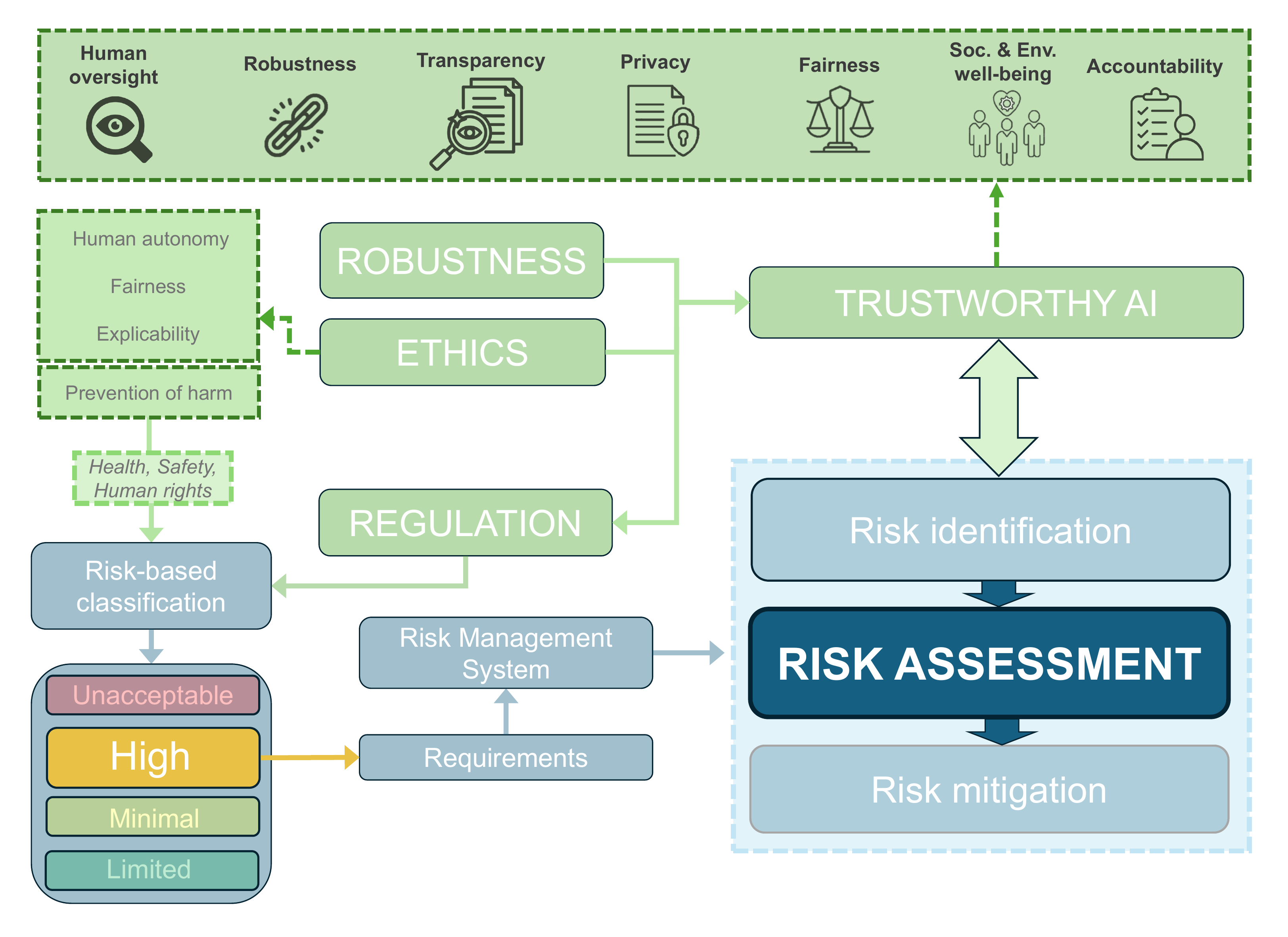}
  \caption{Overview of AI risk assessment as a module in the context of a general responsible AI framework.}

  \label{fig:fig1}
\end{figure*}

Certain clauses within these requirements, notably risk identification, estimation and evaluation of risks from reasonably foreseeable misuse, and evaluation of other possible risks, directly involve \textit{risk assessment} (see Fig.~\ref{fig:fig1}). Such assessments are challenging because AI risks are abstract and multidimensional, and depend on implementation scale, technology, application context, and even the definition of risk itself.

This creates a critical question: how can these multifaceted AI risks be assessed in practice?

\subsection{International Regulatory Summary}

Several jurisdictions have established frameworks that address AI risk assessment, including binding obligations for organizations deploying high-risk systems. The following examples show different approaches to defining, evaluating, and managing AI-associated risks.

\textbf{European Union – Artificial Intelligence Act}: The AI Act\cite{eu2024ai_act} creates a four‐level risk taxonomy and legally obliges providers of \emph{high‑risk} systems—such as Annex III uses in biometric identification, critical infrastructure, hiring, or credit—to conduct extensive risk management. Article 9 requires a documented \textit{risk‑management system} to identify and mitigate risks throughout the AI life cycle. Before deployment or market access in the EU, these systems must pass a conformity assessment covering data quality, transparency, human oversight, robustness, and risk mitigation. Other EU rules also support AI risk assessment in specific contexts; for example, the GDPR requires Data Protection Impact Assessments (DPIAs)\cite{eu2016gdpr} for high-risk personal data processing, including AI-driven profiling or automated decisions.

\textbf{United States – OMB Memorandum M‑24‑10}: While Congress still debates a horizontal AI law, OMB Memorandum M-24-10\cite{omb2024m2410} requires federal agencies to treat AI risks systematically. Systems deemed rights‑impacting (e.g.\ employment or benefits decisions) or safety‑impacting (e.g.\ autonomous vehicles or diagnostics) must undergo pre-deployment risk assessment aligned with the NIST AI RMF. Agencies must map context and stakeholders, measure likelihood and impact, select controls such as bias testing, human oversight and red-team security probes, and continuously monitor performance. A Chief AI Officer coordinates inventories and annual public summaries, while OMB audits compliance. Beyond federal agencies, U.S. mandates remain sector-specific; for instance, the FDA treats some AI-based medical software as a medical device\cite{fda2025ai_samd}, subjecting them to pre-market risk analysis and post-market surveillance.

\textbf{Canada – Directive on Automated Decisions‐Making}: Since 2019, every federal project that automates a decision about people must complete the Online Algorithmic Impact Assessment (AIA)\cite{canada_aia_tool} \emph{before any code ships}. Its 51 risk questions and 34 mitigation questions cover design intent, data sources, algorithmic logic, privacy, bias, transparency, and human fall‑backs. The tool scores systems from Level I to IV: Level I triggers basic documentation, while Level IV (e.g., immigration or parole decisions) demands external peer review, source-code transparency, plain-language explanations, live human override, and yearly re-certification. Published results create public pressure, and the Treasury Board can suspend projects that ignore the AIA.

\textbf{China – Interim Measures on Generative AI Services}: In 2023, regulators introduced rules for commercial generative AI. The Interim Measures for the Administration of Generative AI Services\cite{cac2023genai} impose security evaluation obligations on providers. Article 17 requires services with public opinion or social mobilization capacity to conduct a security evaluation and complete algorithm submission procedures.

\subsubsection{Non-binding official guidelines}

In addition to binding regulations, several countries have developed influential voluntary frameworks that guide responsible AI adoption and risk assessment.

\textbf{Singapore – Model AI Governance Framework (voluntary)}: The framework\cite{pdpc2020ai_framework} urges companies to adopt a risk-based approach: identify AI features with the greatest stakeholder impact, then select proportionate controls such as explainability, robustness testing, auditability, dataset governance, and human-in-the-loop mechanisms. 

\textbf{United Kingdom – AI Regulation White Paper (consultative)}: The white paper\cite{uk2023ai_whitepaper} sets five overarching principles and asks existing regulators to conduct context-specific risk assessments rather than imposing a single statute. A central function will share risk templates, coordinate horizon scanning, and support regulatory sandboxes for proportionate, innovation-friendly oversight. 

\textbf{India – NITI Aayog “Responsible AI for All” (voluntary)}: The roadmap\cite{niti2021responsible_ai_principles} frames seven constitutional principles and calls for calibrated, risk-proportionate assessment, from system-level issues such as bias, explainability, and privacy to wider societal impacts. 

\textbf{United States (industry) – NIST AI RMF 1.0 (voluntary)}: NIST’s framework offers sector-agnostic guidance: organizations Govern AI risk, Map context \& stakeholders, Measure likelihood/impact with metrics, and Manage through controls and monitoring, complemented by a Playbook and sector profiles (e.g., for generative AI).

\textbf{International Standards – ISO/IEC 42001 \& 23894 (certifiable)}: ISO 42001 allows organizations to certify an AI management system covering policy, roles, objectives, and continuous improvement, while ISO 23894 provides techniques to identify, analyze, evaluate, and treat AI-specific risks, aligning with ISO 31000 and NIST AI RMF. 

\subsubsection{AI Risk Management Standards}

International standardization efforts support the practical implementation of AI risk assessment required by regulations. Particularly relevant are ISO and IEC standards, especially within ISO/IEC JTC 1/SC 42. For the AI Act RMS, these include ISO/IEC AWI 5338, TR 5469, 23894-2, 24027, 24029-1, CD 24668, 38507, and 42001\cite{nativi2021ai}. These standards address different aspects of AI risk management and support compliance and best practices.

The National Institute of Standards and Technology (NIST) has developed complementary frameworks for AI risk management, emphasizing flexibility, transparency, and accountability. These frameworks add practical approaches and methodologies for regulators and AI developers.

The NIST AI Risk Management Framework (AI RMF 1.0, NIST.AI.100-1) provides guidance to identify, measure, and prioritize AI risks throughout the AI system life cycle. It quantifies risks by assessing likelihood and impact, using qualitative and, where possible, quantitative metrics to capture AI complexity and uncertainty. Its four functions—GOVERN (oversight and accountability), MAP (contextualizing risks), MEASURE (applying metrics), and MANAGE (implementing treatment strategies)—structure risk analysis and management.

%\begin{table}[h]
%\centering
%\resizebox{\textwidth}{!}{
%\begin{tabular}{|l|l|l|}
%\hline
%\textbf{Trait} & \textbf{Example Metric} & \textbf{How It’s Measured} \\
%\hline
%Accuracy & F1-score, ROC-AUC & Labeled test sets \\
%\hline
%Fairness & Demographic parity & Outcome comparisons across groups \\
%\hline
%Explainability & SHAP/LIME, user trust surveys & Model introspection + human studies \\
%\hline
%Privacy & Membership inference resistance & Attack simulations \\
%\hline
%Robustness & Adversarial error rate & Perturbed input testing \\
%\hline
%Security & Attack recovery time & Fault injection + recovery analysis \\
%\hline
%Transparency & Auditability, completeness & Documentation reviews \\
%\hline
%\end{tabular}
%}
%\caption{Examples of AI Risk Metrics and Their Measurement Approaches}
%\label{tab:ai-metrics}
%\end{table}

The Generative AI Profile (NIST.AI.600-1) extends the AI RMF by addressing risks unique to or exacerbated by Generative AI (GAI), such as confabulation, harmful content generation, privacy concerns, cybersecurity vulnerabilities, and intellectual property issues. It introduces measures such as Content Provenance, Pre-deployment Testing, and Incident Disclosure to improve transparency, responsiveness, empirical testing, and continuous monitoring of generative AI risks.

\section{Risk Assessment: Identification and Analysis}

\subsection{Risk Identification}

%AI act requirements $\rightarrow$ Risk management system $\rightarrow$ Risk management process $\rightarrow$ 1) identification of risks associated with AI system, 2) estimation and evaluation of the risks caused by (reasonably) foreseeable misuse and 3) evaluation of other possible risks \cite{nativi2021ai}

%AI Ethic guidelines for trustworthy AI: 7 principles \footnote{https://digital-strategy.ec.europa.eu/en/library/ethics-guidelines-trustworthy-ai}

%Standards: NIST, ISO(ISO/IEC AWI 5259-3, ISO/IEC AWI 5338, TR 5469, 23894-2, 24027, 24029-1, CD 24668, 38507, 42001)

AI risk assessment involves two stages: risk identification, which pinpoints specific hazards or sources of harm, and risk analysis, which evaluates their likelihood and severity \cite{steimers2022sources}. Effective identification requires a structured framework for mapping potential risks, but since AI systems are complex and "risk" is an abstract concept, researchers have proposed varied taxonomies, each capturing different technical and ethical challenges.
This matters more than ever as AI spreads into high-stakes sectors like healthcare, finance, transportation, and security. A key step toward good governance is understanding these different risk types, which is why many researchers have developed frameworks, methodologies, and classifications to help identify, prioritize, and mitigate AI-related risks.

A critical commonality among these frameworks is the categorization of AI risks into technical and non-technical dimensions. Bagehorn et al.\cite{bagehorn2025ai}, for instance, develop the "AI Risk Atlas" segmenting risks into input, inference, output, and non-technical categories, each further analyzed through dimensions such as accuracy, fairness, privacy, robustness, and explainability. Slattery et al.\cite{slattery2024ai} complement this perspective by synthesizing existing classifications into an accessible "AI Risk Repository," which offers a dual taxonomy: one that categorizes risks causally—by entity, intentionality, and timing—and another by domain, including misinformation, discrimination, privacy, and system security.

Identifying risks specific to particular social-scale contexts also emerges as a common theme in the literature. Critch and Russell\cite{critch2023tasra} highlight the social-scale risks originating from intentional misuse, unintended interactions, and system misalignments, using fault tree analysis to systematically explore accountability and intervention strategies. Similarly, Uuk et al.\cite{uuk2024taxonomy} emphasize the systemic nature of the risks arising from general-purpose AI, pinpointing broad societal threats such as democratic erosion, economic disruptions, and environmental harm. Both studies advocate for structured and comprehensive policy responses to manage these widespread impacts effectively.

The complexity and rapid evolution of AI technologies require adaptive and context-sensitive regulatory frameworks, as evidenced by Al-Maamari et al.\cite{al2025between}. Their comparative analysis across regions including the EU, US, UK, and China underscores how different governance models—ranging from centralized directives to sector-specific approaches—address the delicate balance between innovation and oversight. Each model has strengths and limitations; for example, the structured transparency and conformity assessments of the EU contrast with the decentralized approach of the US, which promotes innovation but risks fragmented enforcement.

Addressing practical implementation and ongoing management of these frameworks, Habbal et al.\cite{habbal2024artificial} introduce the AI Trust, Risk, and Security Management (AI TRiSM) framework, stressing the importance of adaptive strategies such as ModelOps to deal with continuously emerging threats and challenges such as adversarial attacks, biases, and ethical concerns. Steimers and Schneider\cite{steimers2022sources} further reinforce the need for AI-specific risk management strategies, highlighting the unique risks of AI systems, including unpredictability and opacity, that traditional software risk management does not adequately address.

Then, Hendrycks et al.\cite{hendrycks2023overview} offer a perspective on catastrophic risks, categorizing them as malicious use, AI race conditions, organizational risks, and rogue AI scenarios. They propose concrete methods to mitigate these threats, such as enhanced biosecurity, strict model access controls, organizational risk culture improvements, and international regulatory coordination, reflecting the multifaceted approach needed to protect against severe outcomes.

%\begin{landscape}  % pdflscape environment
\begin{table}[t]
  \centering
  \tiny
  \caption{Taxonomies of AI Risks}
  \label{tab:your_label}
%  \begin{adjustbox}{max width=\linewidth, max totalheight=\textheight, keepaspectratio}
  \begin{tabular}{|c|c|c|}
    \Xhline{1.5pt}
    \textbf{Ref.} & \textbf{Taxonomy} & \textbf{Risk category}  \\
    \Xhline{1.5pt}

    % Slattery
    \multirow{10}{*}{Slattery et al.\cite{slattery2024ai}} & \multirow{3}{*}{Causal (high-level)} & Entity \\
                           &                          & Intentionality  \\
                           &                          & Timing  \\
    \cline{2-3}
                           & \multirow{7}{*}{Domain (mid-level)} & Discrimination and toxicity  \\
                           &                          & Privacy and security  \\
                           &                          & Misinformation  \\
                           &                          & Malicious actors and misuse  \\
                           &                          & Human-computer interaction  \\
                           &                          & Socioeconomic and environmental  \\
                           &                          & AI safety, failures and limitations \\
    \Xhline{1.5pt}

    % Uuk
    \multirow{13}{*}{Uuk et al.\cite{uuk2024taxonomy}} & \multirow{13}{*}{Systemic} & Control \\
                           &                          & Democracy  \\
                           &                          & Discrimination  \\
                           &                          & Economy  \\
                           &                          & Environment  \\
                           &                          & Fundamental rights  \\
                           &                          & Governance  \\
                           &                          & Harms to non-humans  \\
                           &                          & Information  \\
                           &                          & Irreversible change  \\
                           &                          & Power  \\
                           &                          & Security  \\
                           &                          & Warfare  \\
    \Xhline{1.5pt}

    % Steimers
    \multirow{8}{*}{Steimers et al.\cite{steimers2022sources}} & \multirow{3}{*}{Ethical} & Fairness  \\
                           &                          & Privacy  \\
                           &                          & Degree of automation  \\
    \cline{2-3}
                           & \multirow{5}{*}{Reliability and robustness} & Complexity of the task and usage environment  \\
                           &                          & Degree of transparency and explainability \\
                           &                          & Security  \\
                           &                          & System hardware \\
                           &                          & Technological maturity  \\
    \Xhline{1.5pt}
    % OECD
    \multirow{5}{*}{OECD\cite{oecd0449}} & \multirow{5}{*}{Trustworthy AI} & Sustainability \\
                           &                          & Human rights, privacy and fairness  \\
                           &                          & Transparency and explainability \\
                           &                          & Robustness, security and safety \\
                           &                          & Accountability  \\

%    \Xhline{1.5pt}
%    \multicolumn{10}{|p{\linewidth}|}{\small 
%    \textbf{Legend:} H: Human Agency and Oversight; R: Technical Robustness and Safety; P: Privacy and Data Governance; T: Transparency; F: Diversity, Non-discrimination, and Fairness; E: Societal and Environmental Well-being; A: Accountability
%    } \\
    \Xhline{1.5pt}
  \end{tabular}
\end{table}
%\end{landscape}
  \normalsize

\subsection{Risk Analysis}

Different studies have proposed various approaches to measure and manage risks derived from the use of AI, highlighting a diverse range of frameworks and conceptualizations. This diversity underscores the complexity and richness of the landscape, showcasing non-exclusive, complementary approaches to AI risk assessment.

The fundamental need for clear governance and structured frameworks to manage AI risks is addressed comprehensively by Falco et al.\cite{falco2021governing}, who propose an independent audit system known as IAAIS. They emphasize prospective assessments, audit trails, and adherence to jurisdictional requirements, ensuring operational transparency and fostering public trust. Similarly, Koshiyama et al.\cite{koshiyama2022algorithm} underscore the systemic necessity of rigorous auditing, highlighting how comprehensive lifecycle assessments can mitigate legal, ethical, and operational risks through specialized auditing processes.

Moving from governance to implementation, Felländer et al.\cite{fellander2022achieving} introduce DRESS-eAI, a novel data-driven methodology. Their approach integrates multidisciplinary perspectives—including technical, legal, and social insights—to address ethical risks practically and sustainably. This aligns closely with Nagbøl et al.\cite{nagbol2021designing}, whose AIRA tool specifically prioritizes structured stakeholder communication and broad performance metrics, such as fairness, interpretability, and privacy, promoting effective organizational risk management.

Further developing these ideas, Giudici et al. \cite{giudici2024artificial} present the Key AI Risk Indicators (KAIRI) framework, directly tying regulatory compliance from the European AI Act to quantifiable principles like Sustainability, Accuracy, Fairness and Explainability, particularly in financial contexts. Meanwhile, Zhang et al. \cite{zhang2022towards} broaden this approach by categorizing risks at the data and model levels, highlighting the critical importance of managing biases, uncertainties, and potential adversarial vulnerabilities in high-risk AI applications. Such domain-oriented operationalizations are also emerging in education, where Irigoyen et al. \cite{irigoyen2026edueval,irigoyen2026AIrisks-edu} frame pedagogical risk evaluation through an extended dataset for explainable assessment, and in human--AI interaction, where Daza et al. \cite{daza2026ares} assess social-engineering risks through adaptive role-based evaluation.

Based on regulatory frameworks, Novelli et al. \cite{novelli2024ai} advocate for proportional and scenario-based risk assessment models inspired by climate change risk frameworks. Their methodology examines complex interactions among risk determinants, enhancing the precision and adaptability of regulatory responses, particularly relevant for general-purpose AI systems such as large language models (LLMs).

To bridge theory with practical application, Koessler et al.\cite{koessler2023risk} review established risk assessment methods from safety-critical industries, recommending their adaptation for AI-specific catastrophic risks. Techniques like scenario analysis, causal mapping, and Delphi methods are promoted for iterative pre-deployment and pre-training assessments, ensuring comprehensive life cycle coverage. Complementing this, Xia et al.\cite{xia2023towards} provide an extensive mapping of current frameworks, highlighting strengths, limitations, and recommending enhancements towards concreteness and interconnectedness in future AI risk assessments.

Collectively, these contributions form a holistic and dynamic understanding of AI risk assessment, advocating robust interdisciplinary frameworks, precise regulatory alignment, and structured stakeholder engagement to navigate the complex risk landscape effectively.

\section{Future Work}

In our future work, we will study risks in multimodal LLMs (including VLMs \cite{dealcala2026demo2}) and representations based on images of agents (including avatars~\cite{laura2026ava}). Analyzing biases \cite{2023_ECAIw_LFIT-XAI_Tello,pena2025addressing} and synthetic manipulations \cite{pavel25iccv} while maintaining privacy \cite{mancera2025pba} are also key topics for risk evaluation on our agenda. Finally, we will apply these risk assessment and management principles to concrete domains such as document intelligence \cite{MUNOZHARO2026103969}, e-learning~\cite{irigoyen2026AIrisks-edu,becerra2024biometrics}, gaming~\cite{daza2026ares}, and e-health~\cite{ROMEROTAPIADOR2026111676}.

\bibliographystyle{IEEEtran}
\bibliography{IEEEabrv,bibliography}

\begin{thebibliography}{10}
\providecommand{\url}[1]{#1}
\csname url@samestyle\endcsname
\providecommand{\newblock}{\relax}
\providecommand{\bibinfo}[2]{#2}
\providecommand{\BIBentrySTDinterwordspacing}{\spaceskip=0pt\relax}
\providecommand{\BIBentryALTinterwordstretchfactor}{4}
\providecommand{\BIBentryALTinterwordspacing}{\spaceskip=\fontdimen2\font plus
\BIBentryALTinterwordstretchfactor\fontdimen3\font minus \fontdimen4\font\relax}
\providecommand{\BIBforeignlanguage}[2]{{%
\expandafter\ifx\csname l@#1\endcsname\relax
\typeout{** WARNING: IEEEtran.bst: No hyphenation pattern has been}%
\typeout{** loaded for the language `#1'. Using the pattern for}%
\typeout{** the default language instead.}%
\else
\language=\csname l@#1\endcsname
\fi
#2}}
\providecommand{\BIBdecl}{\relax}
\BIBdecl

\bibitem{cannarsa2021ethics}
M.~Cannarsa, ``Ethics guidelines for trustworthy {AI},'' \emph{The Cambridge Handbook of Lawyering in the Digital Age}, pp. 283--297, 2021.

\bibitem{2023_SNCS_Human-Centric_Pena}
A.~Peña \emph{et~al.}, ``Human-centric multimodal machine learning: Recent advances and testbed on {AI}-based recruitment,'' \emph{SN Computer Science}, vol.~4, no.~5, p. 434, June 2023.

\bibitem{diaz2023connecting}
N.~D{\'\i}az-Rodr{\'\i}guez, J.~Del~Ser, M.~Coeckelbergh, M.~L. de~Prado, E.~Herrera-Viedma, and F.~Herrera, ``Connecting the dots in trustworthy artificial intelligence: From {AI} principles, ethics, and key requirements to responsible {AI} systems and regulation,'' \emph{Information Fusion}, vol.~99, p. 101896, 2023.

\bibitem{eu2024ai_act}
\BIBentryALTinterwordspacing
{European Parliament and Council of the European Union}, ``{Regulation (EU) 2024/1689 of 13 June 2024 laying down harmonised rules on artificial intelligence (Artificial Intelligence Act)},'' Official Journal of the European Union, L 2024/1689, Jul. 2024. [Online]. Available: \url{https://eur-lex.europa.eu/eli/reg/2024/1689/oj/eng}
\BIBentrySTDinterwordspacing

\bibitem{eu2016gdpr}
{European Parliament and Council of the European Union\relax}, ``{General Data Protection Regulation (GDPR)},'' {Regulation (EU) 2016/679, Official Journal of the European Union, L 119}, Apr. 2016, [Online]. Available: \href{https://eur-lex.europa.eu/eli/reg/2016/679/oj/eng}{EUR-Lex}.

\bibitem{omb2024m2410}
{Office of Management and Budget}, ``{Advancing Governance, Innovation, and Risk Management for Agency Use of Artificial Intelligence},'' {Memorandum M-24-10, Executive Office of the President}, Mar. 2024, [Online]. Available: \href{https://www.whitehouse.gov/wp-content/uploads/2024/03/M-24-10-Advancing-Governance-Innovation-and-Risk-Management-for-Agency-Use-of-Artificial-Intelligence.pdf}{White House PDF}.

\bibitem{fda2025ai_samd}
{U.S. Food and Drug Administration\relax}, ``{Artificial Intelligence in Software as a Medical Device},'' Mar. 2025, [Online]. Available: \href{https://www.fda.gov/medical-devices/software-medical-device-samd/artificial-intelligence-software-medical-device}{FDA AI SaMD page}.

\bibitem{canada_aia_tool}
\BIBentryALTinterwordspacing
{Government of Canada}, ``{Algorithmic Impact Assessment Tool},'' 2025. [Online]. Available: \url{https://www.canada.ca/en/government/system/digital-government/digital-government-innovations/responsible-use-ai/algorithmic-impact-assessment.html}
\BIBentrySTDinterwordspacing

\bibitem{cac2023genai}
\BIBentryALTinterwordspacing
{CAC et al.}, ``{Interim Measures for the Administration of Generative Artificial Intelligence Services},'' Jul. 2023, order No. 15. [Online]. Available: \url{https://www.cac.gov.cn/2023-07/13/c_1690898327029107.htm}
\BIBentrySTDinterwordspacing

\bibitem{pdpc2020ai_framework}
\BIBentryALTinterwordspacing
{PDPC Singapore and IMDA}, ``{Model Artificial Intelligence Governance Framework},'' {PDPC Singapore and IMDA}, Tech. Rep., Jan. 2020. [Online]. Available: \url{https://www.pdpc.gov.sg/-/media/files/pdpc/pdf-files/resource-for-organisation/ai/sgmodelaigovframework2.pdf}
\BIBentrySTDinterwordspacing

\bibitem{uk2023ai_whitepaper}
\BIBentryALTinterwordspacing
{DSIT}, ``{A Pro-Innovation Approach to AI Regulation},'' {HM Government}, Tech. Rep. CP 815, Mar. 2023. [Online]. Available: \url{https://www.gov.uk/government/publications/ai-regulation-a-pro-innovation-approach/white-paper}
\BIBentrySTDinterwordspacing

\bibitem{niti2021responsible_ai_principles}
\BIBentryALTinterwordspacing
{NITI Aayog}, ``{Responsible AI \#AIForAll: Approach Document for India, Part 1 -- Principles for Responsible AI},'' {NITI Aayog}, Tech. Rep., Feb. 2021. [Online]. Available: \url{https://www.niti.gov.in/sites/default/files/2021-02/Responsible-AI-22022021.pdf}
\BIBentrySTDinterwordspacing

\bibitem{nativi2021ai}
S.~Nativi and D.~Nigris, ``{AI watch: AI standardisation landscape},'' \emph{European Commission}, 2021.

\bibitem{steimers2022sources}
A.~Steimers and M.~Schneider, ``Sources of risk of {AI} systems,'' \emph{Intl. Journal of Environmental Research and Public Health}, 2022.

\bibitem{bagehorn2025ai}
F.~Bagehorn, K.~Brimijoin, E.~M. Daly, J.~He, M.~Hind, L.~Garces-Erice, C.~Giblin, I.~Giurgiu, J.~Martino, R.~Nair \emph{et~al.}, ``{AI} risk atlas: Taxonomy and tooling for navigating {AI} risks and resources,'' \emph{arXiv preprint arXiv:2503.05780}, 2025.

\bibitem{slattery2024ai}
P.~Slattery, A.~K. Saeri, E.~A. Grundy, J.~Graham \emph{et~al.}, ``The {AI} risk repository: A comprehensive meta-review, database, and taxonomy of risks from artificial intelligence,'' \emph{arXiv:2408.12622}, 2024.

\bibitem{critch2023tasra}
A.~Critch and S.~Russell, ``{TASRA}: a taxonomy and analysis of societal-scale risks from {AI},'' \emph{arXiv preprint arXiv:2306.06924}, 2023.

\bibitem{uuk2024taxonomy}
R.~Uuk, C.~I. Gutierrez \emph{et~al.}, ``A taxonomy of systemic risks from general-purpose {AI},'' \emph{arXiv:2412.07780}, 2024.

\bibitem{al2025between}
A.~Al-Maamari, ``Between innovation and oversight: A cross-regional study of {AI} risk management frameworks in the {EU, US, UK, and China},'' \emph{arXiv preprint arXiv:2503.05773}, 2025.

\bibitem{habbal2024artificial}
A.~Habbal, M.~K. Ali, and M.~A. Abuzaraida, ``Artificial intelligence trust, risk and security management ({AI} trism): Frameworks, applications, challenges and future research directions,'' \emph{Expert Systems with Applications}, vol. 240, p. 122442, 2024.

\bibitem{hendrycks2023overview}
D.~Hendrycks, M.~Mazeika, and T.~Woodside, ``An overview of catastrophic {AI} risks,'' \emph{arXiv preprint arXiv:2306.12001}, 2023.

\bibitem{oecd0449}
\BIBentryALTinterwordspacing
``{OECD} legal instrument 0449.'' [Online]. Available: \url{https://legalinstruments.oecd.org/en/instruments/OECD-LEGAL-0449}
\BIBentrySTDinterwordspacing

\bibitem{falco2021governing}
G.~Falco \emph{et~al.}, ``Governing {AI} safety through independent audits,'' \emph{Nature Machine Intelligence}, vol.~3, no.~7, pp. 566--571, 2021.

\bibitem{koshiyama2022algorithm}
A.~Koshiyama \emph{et~al.}, ``Algorithm auditing: Managing the legal, ethical, and technological risks of artificial intelligence, machine learning, and associated algorithms,'' \emph{Computer}, vol.~55, no.~4, pp. 40--50, 2022.

\bibitem{fellander2022achieving}
A.~Fell{\"a}nder, J.~Rebane \emph{et~al.}, ``Achieving a data-driven risk assessment methodology for ethical {AI},'' \emph{Digital Society}, vol.~1, no.~2, p.~13, 2022.

\bibitem{nagbol2021designing}
P.~R. Nagb{\o}l, O.~M{\"u}ller, and O.~Krancher, ``Designing a risk assessment tool for artificial intelligence systems,'' in \emph{Intl. Conf. on Design Science Research in Information Systems and Technology}, 2021, pp. 328--339.

\bibitem{giudici2024artificial}
P.~Giudici, M.~Centurelli \emph{et~al.}, ``Artificial intelligence risk measurement,'' \emph{Expert Systems with Applications}, vol. 235, p. 121220, 2024.

\bibitem{zhang2022towards}
X.~Zhang, F.~T. Chan, C.~Yan, and I.~Bose, ``Towards risk-aware artificial intelligence and machine learning systems: An overview,'' \emph{Decision Support Systems}, vol. 159, p. 113800, 2022.

\bibitem{irigoyen2026edueval}
J.~Irigoyen, R.~Daza \emph{et~al.}, ``{EduEVAL-DB: A Role-Based Dataset for Pedagogical Risk Evaluation in Educational Explanations},'' in \emph{Int. Conf. on Learning Analytics \& Knowledge Workshops (GenAI-LA)}, 2026.

\bibitem{irigoyen2026AIrisks-edu}
J.~Irigoyen, R.~Daza, F.~Jurado, J.~Fierrez, R.~Tolosana, A.~Ortigosa \emph{et~al.}, ``{{AIriskEval-edu: New} Dataset for Risk Assessment in {AI-mediated K-12} Educational Explanations},'' in \emph{IEEE ICCST}, 2026.

\bibitem{daza2026ares}
R.~Daza \emph{et~al.}, ``{Evaluating Social Engineering Risks in AI-based Interaction using Biometrics and a Gaming Setup},'' in \emph{ICCST}, 2026.

\bibitem{novelli2024ai}
C.~Novelli, F.~Casolari, A.~Rotolo, M.~Taddeo, and L.~Floridi, ``{AI} risk assessment: A scenario-based, proportional methodology for the {AI Act},'' \emph{Digital Society}, vol.~3, no.~1, p.~13, 2024.

\bibitem{koessler2023risk}
L.~Koessler and J.~Schuett, ``Risk assessment at {AGI} companies: A review of popular risk assessment techniques from other safety-critical industries,'' \emph{arXiv preprint arXiv:2307.08823}, 2023.

\bibitem{xia2023towards}
B.~Xia \emph{et~al.}, ``Towards concrete and connected {AI} risk assessment ({C2AIRA}): A systematic mapping study,'' in \emph{Intl. Conf. on AI Engineering (CAIN)}, 2023, pp. 104--116.

\bibitem{dealcala2026demo2}
D.~DeAlcala \emph{et~al.}, ``Is my vision-language data in your {AI}? membership inference test ({MINT}) {D}emo 2,'' in \emph{IEEE COMPSAC}, 2026.

\bibitem{laura2026ava}
L.~Pedrouzo \emph{et~al.}, ``Leveraging avatar fingerprinting: A photorealistic talking-head public database and benchmark,'' \emph{arXiv:2603.26934}, 2026.

\bibitem{2023_ECAIw_LFIT-XAI_Tello}
J.~Tello, M.~de~la Cruz, T.~Ribeiro \emph{et~al.}, ``Symbolic {AI (LFIT) for XAI} to handle biases,'' in \emph{European Conf. on Artificial Intelligence Workshops (ECAIw)}, ser. CEUR-WS, vol. 3523, October 2023.

\bibitem{pena2025addressing}
A.~Pe{\~n}a \emph{et~al.}, ``Addressing bias in {LLMs}: Strategies and application to fair {AI}-based recruitment,'' in \emph{AAAI/ACM AIES}, 2025.

\bibitem{pavel25iccv}
P.~Korshunov \emph{et~al.}, ``{DeepID} challenge of detecting synthetic manipulations in {ID} documents,'' in \emph{IEEE ICCV Workshops}, 2025.

\bibitem{mancera2025pba}
G.~Mancera \emph{et~al.}, ``{PBa-LLM: Privacy-and bias-aware NLP using named-entity recognition (NER)},'' in \emph{IAPR ICDAR}.\hskip 1em plus 0.5em minus 0.4em\relax Springer, 2025.

\bibitem{MUNOZHARO2026103969}
J.~Muñoz-Haro, R.~Tolosana \emph{et~al.}, ``Privacy-aware detection of fake identity documents: methodology, benchmark, and improved algorithms ({FakeIDet2}),'' \emph{Information Fusion}, vol. 128, p. 103969, 2026.

\bibitem{becerra2024biometrics}
{\'A}.~Becerra, J.~Irigoyen, R.~Daza, R.~Cobos, A.~Morales, J.~Fierrez, and M.~Cukurova, ``Biometrics and behavior analysis for detecting distractions in e-learning,'' in \emph{2024 International Symposium on Computers in Education (SIIE)}.\hskip 1em plus 0.5em minus 0.4em\relax IEEE, 2024, pp. 1--6.

\bibitem{ROMEROTAPIADOR2026111676}
S.~Romero-Tapiador, R.~Tolosana, A.~Morales \emph{et~al.}, ``Personalized weight loss management through wearable devices and artificial intelligence,'' \emph{Computers in Biology and Medicine}, vol. 209, p. 111676, 2026.

\end{thebibliography}

\end{document}